\documentclass[a4paper,fleqn,usenatbib]{mnras}

\usepackage[T1]{fontenc}
\usepackage{ae,aecompl}

\usepackage{graphicx}	
\usepackage{amsmath,subeqnarray}	
\usepackage{amssymb}	
\usepackage{epsfig}

\title[Role of Non-ionizing radiation]{The Role of Non-ionizing Radiation Pressure in Star Formation: The Stability of Cores and Filaments}

\author[Seo \& Youdin]{
Young Min Seo,$^{1}$\thanks{E-mail: seo3919@email.arizona.com}
Andrew N. Youdin$^{1}$
\\
$^{1}$Department of Astronomy \& Steward Observatory, University of Arizona, 933 N. Cherry Ave., Tucson, AZ 85721, USA\\
}

\date{Accepted XXX. Received YYY; in original form ZZZ}

\pubyear{2016}

\begin{document}
\label{firstpage}
\pagerange{\pageref{firstpage}--\pageref{lastpage}}
\maketitle

\begin{abstract}
Stars form when filaments and dense cores in molecular clouds fragment and collapse due to self-gravity.
In the most basic analyses of gravitational stability, the competition between self-gravity and thermal pressure sets the critical (i.e. maximum stable) mass of  spheres and the critical line density of cylinders.  Previous work has considered additional support from magnetic fields and turbulence.  Here, we consider the effects of non-ionizing radiation, specifically the inward radiation pressure force that acts on dense structures embedded in an isotropic radiation field.  Using hydrostatic, isothermal models, we find that irradiation lowers the critical mass and line density for gravitational collapse, and can thus act as a trigger for star formation.  For structures with moderate central densities, $\sim10^3$ cm$^{-3}$, the interstellar radiation field in the Solar vicinity has an order unity effect on stability thresholds.   For more evolved objects with higher central densities, a significant lowering of stability thresholds requires stronger irradiation, as can be found closer to the Galactic center or near stellar associations.  Even when strong sources of ionizing radiation are absent or extincted, our study shows that interstellar irradiation can significantly influence the star formation process.

\end{abstract}




\section{INTRODUCTION}

Star formation is a process by which interstellar gas becomes denser via a hierarchy of structures, with gravitational collapse playing a key role.   Dense filaments are ubiquitous in molecular clouds, with  75\% of denser starless cores residing in filaments \citep{andre10}. The formation of cores within filaments is explained by  gravitational instability \citep{inutsuka92}.  The formation of stars within cores is attributed to another gravitational collapse \citep{sal87}.

A basic understanding of gravitational collapse comes from the study of isothermal and pressure-confined gas in hydrostatic equilibrium.  The classic solutions are the Bonnor-Ebert (BE) sphere  \citep{ebert55,ebert57,bonnor56} and the isothermal  \citet{ostriker64} cylinder.
These studies show that gravitational collapse occurs above the critical mass of the BE sphere and the critical line density (mass per length) of isothermal cylinders.  The critical BE mass also applies to hydrostatic clouds of any geometry, provided the volume is finite \citep{lombardi01}.    Observed density profiles of filaments and cores are often well matched by these simple models \citep[e.g.][]{bacmann00, alves01, kandori05, hacar11}.

While of fundamental importance, these classic solutions neglect many potentially significant effects.  Magnetic fields, turbulence and detailed radiative transfer can alter the structure and stability of cores and filaments \citep{mckee07}.
This work focuses on a particular aspect of radiative transfer, the radiation pressure exerted on cores and filaments by ambient non-ionizing radiation.  Ionizing radiation is known to have important effects on star formation in HII regions, i.e.\ near high mass stars.  Our focus on non-ionizing radiation applies not only to low mass star forming regions, but also to regions of high column density into which non-ionizing radiation penetrates more deeply.

Non-ionizing radiation (from mid-UV to mid-IR) exerts radiation pressure on dust grains which are frictionally coupled to the gas \citep{draine11}.   The radiation pressure force is weak in the diffuse interstellar medium because the interstellar radiation field (hereafter ISRF) is almost isotropic, with $\sim$10\% anisotropy, \citep{weingartner01}.
However, near and inside a dense structure, radiation becomes anisotropic due to shadowing by the structure itself.  With this introduced anisotropy, radiation pressure becomes comparable to thermal gas pressure in the interstellar medium.

In this paper we study how the radiation pressure force alters the structure and gravitational stability of filaments and cores.
Section \ref{sec:model} describes our model for hydrostatic irradiated structures.
Section \ref{sec:selfsim} presents our self-similar solutions in dimensionless coordinates.   In \S4, we apply our results to the physical conditions of star forming regions. In \S5, we discuss limitations and future extensions of our model. We summarize our results in \S6.

\section{Model for Irradiated Cylinders and Spheres}\label{sec:model}

\subsection{Hydrostatic Structure with Radiation Pressure}

We consider hydrostatic configurations of dense cores and filaments exposed to non-ionizing photons that exert radiation pressure. Our idealized model assumes spherical symmetry for dense cores and cylindrical symmetry for filaments. Radiation pressure acts on dust grains, which are uniformly mixed and perfectly coupled to the gas. Possible sedimentation of dust grains is addressed in \S5.

Our models satisfy hydrostatic equilibrium and the Poisson equation:
\begin{align}
&\nabla P_{\rm g} = -\rho\nabla\psi+n_{\rm d}\vec{f}_{\rm rad}  \label{Hbal} \\
&\nabla^2\psi = 4\pi G \rho  \label{poisson}
\end{align}
where $P_{\rm g}$ is the gas pressure, $\psi$ is the gravitational potential, $\vec{f}_{\rm rad}$ is the force exerted to a dust grain due to radiation pressure, $n_{\rm d}$ is the number density of dust grains per unit volume, and $\rho$  = $\rho_{\rm g}$ + $\rho_{\rm d}$ is the total density of dust and gas, respectively.

The radiation pressure force on a dust grain is
\begin{equation}
\vec{f}_{\rm rad}~=\pi a^2 \langle Q_{\rm pr}\rangle {\vec{F}_{\rm rad} \over c},
\label{rad}
\end{equation}
where $a$ is the effective spherical radius of dust grains, $\vec{F}_{\rm rad}$ is the energy flux of radiation field,  $c$ is the speed of light and $\langle Q_{\rm pr}\rangle$ is radiation pressure efficiency, described in detail in \S4.1.  The mass of a dust grain is
\begin{equation}
m_{\rm d} = {\rho_{\rm d} \over n_{\rm d}} = {4\pi \over 3 }  \rho_{\rm m} a^3
\label{nd}\end{equation}
with $\rho_{\rm m}$ the internal, material density of dust grains.

\begin{figure}
  \begin{center}
    \epsfxsize = 3.in
    \epsffile{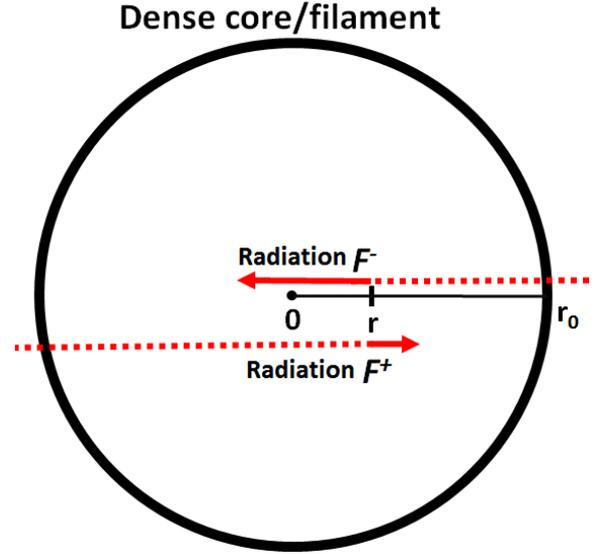}
  \end{center}
\caption{Schematic of our two-ray model for the flux of non-ionizing radiation inside a dense core or filament. The net flux at any internal radius $r < r_0$ arises from two competing rays: the inward directed flux, $F^-$, and the outward directed flux, $F^+$, that has passed though the center of the object.  An optically thin object in an isotropic radiation field will experience little net flux as the two contributions nearly cancel.}
\label{fig:two_ray}
\end{figure}

\begin{figure*}
  \begin{center}
    \leavevmode
    \epsfxsize = 6.5in
    \epsffile{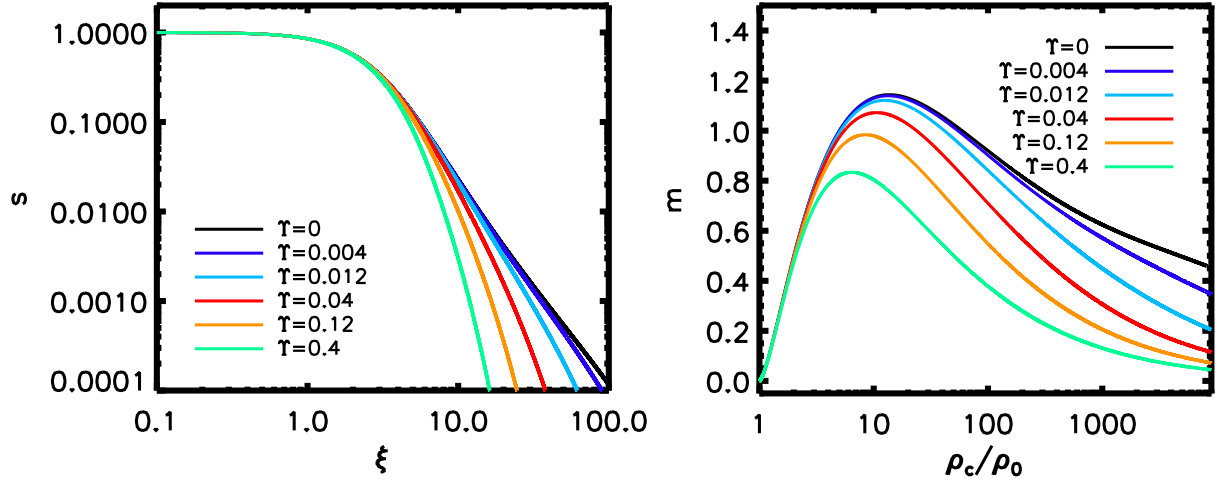}
  \end{center}
\caption{ ({\it Left}): Dimensionless profiles of density, $s$, vs.\ radius, $\xi$, for spheres subject to different levels of (non-ionizing) irradiation, $\Upsilon$, as labelled for the colored curves.  ({\it Right}): Dimensionless mass, $m$, as a function of density contrast, where maximum values correspond to marginal gravitational stability, as described in the text. The dimensionless dust opacity is fixed to $\zeta = 2.6$ in both plots.}
\label{profile_core}
\end{figure*}

The radiative flux is directed inwards, $\vec{F}_{\rm rad} = F \hat{r}$ with $F<0$ and $\hat{r}$ the  unit vector along the spherical or cylindrical radial coordinate, $r$.  We consider the extinction of radiation (but ignore scattering) by taking a two-ray approximation (see Figure \ref{fig:two_ray}). One ray, $F^-$, represents the flux entering the sphere/cylinder from the near surface, while $F^+$ represents the oppositely directed flux from the far surface, which has passed through the center. The total flux $F = -F^- + F^+$ with
\begin{subeqnarray}\label{Fs}
F^-&=& F_\circ \exp[-\tau(r) ] \slabel{F-} \\
F^+&=& F_\circ \exp[-\tau_{\rm tot}+\tau(r) ], \label{F+}\, .
\end{subeqnarray}
The optical depth to the near surface, at $r = r_0$, is
\begin{equation}
\tau(r)=\pi \int_r^{r_0} a^2\langle Q_{\rm ext}\rangle  n_d dr',
\label{eq:taur}
\end{equation}
$\langle Q_{\rm ext}\rangle$ is the spectrum-averaged extinction coefficient. The total extinction of a sphere/cylinder is  $\tau_{\rm tot} = \tau(-{r_0})$.  The normalization $F_\circ$ gives the one-sided surface flux, i.e.\   $F(r_0) \rightarrow -F_\circ$ for an opaque object with $\tau_{\rm tot} \gg 1$.  The validity of the two-ray approximation is addressed in \S5.

We adopt an isothermal equation of state for the gas:
\begin{equation}
P_{\rm g} = c_{\rm s}^2\rho_{\rm g},
\label{pg}
\end{equation}
where $c_{\rm s}$ is the sound speed of gas, and $\rho_{\rm g}$ is the gas density, consistent with the standard Bonnor-Ebert problem and in at least rough agreement with observed cores and filaments  \citep{evans01, stamatellos07, seo15}.

To derive our version of the Lane-Emden equation, we take the divergence of ($1/\rho$ times) the equation (\ref{Hbal}), which combined with equation (\ref{poisson}) gives
\begin{equation}  \label{eq:LEdim}
c_{\rm s}^2 \nabla \cdot \left( {1 \over \rho} \nabla {\rho \over 1\!+\!Z}\right) =  -4\pi G \rho + \nabla \cdot \left( {Z \over 1\!+\!Z} {\vec{f}_{\rm rad} \over m_{\rm d}}\right)
\end{equation}
where the dust-to-gas ratio, $Z = \rho_{\rm d}/\rho_{\rm g}$. Solution of  equation (\ref{eq:LEdim}) also requires equations (\ref{rad}), (\ref{Fs}) and (\ref{eq:taur}) to specify the radiation force.  To simplify the solution space, we also hold spatially constant both $Z$ and the dust opacities
\begin{equation}
\kappa_i = {Z \over 1+Z} {3 \langle Q_i\rangle \over 4\rho_{\rm m} a}
\end{equation}
where the index $i = $ ``ext'' and ``pr'' labels the radiation pressure and extinction cases.  In practice, we obtain solutions using the dimensionless equations described below.

\subsection{Dimensionless Equations}\label{sec:dimless}
Our dimensionless equations use the central density $\rho_{\rm c}$, the dust-weighted sound speed $c'_{\rm s} = c_{\rm s}/\sqrt{1\!+\!Z}$,  and the characteristic scale height
\begin{equation}
\alpha \equiv  {c'_{\rm s}\over \sqrt{4\pi G \rho_{\rm c}} }
\end{equation}
as scale factors.  The dimensionless variables
\begin{align}
&\xi \equiv  {r \over \alpha}, \label{xi}\\
&s \equiv {\rho \over \rho_{\rm c}},
\label{s}
\end{align}
describe radial distance and density, while $\tau$ is already dimensionless.

The inclusion of radiative effects adds two new dimensionless parameters. The dimensionless extinction,
\begin{equation}
\zeta\equiv \kappa_{\rm ext} \alpha \rho_{\rm c} ,
\label{zeta}
\end{equation}
gives a characteristic (but not the actual) the optical depth.  The dimensionless radiation pressure strength,
\begin{equation}
\Upsilon\equiv \frac{ \langle Q_{\rm pr}\rangle } {\langle Q_{\rm ext}\rangle } { F_\circ/c \over P_{\rm c}}.
\label{upsilon}
\end{equation}
normalizes the radiation pressure to the central pressure, $P_{\rm c} = {c'_{\rm s}}^2 \rho_{\rm c}$, times the ratio of radiative efficiencies.

In these dimensionless units, the governing equations (\ref{eq:taur}) and (\ref{eq:LEdim}) read
\begin{align}
&{d \tau \over d \xi} = \zeta s , \label{eq:taux}\\
&\nabla_\xi^2 \ln(s) = -s + \zeta \Upsilon A(\xi; \tau) \label{eq:hydros}
\end{align}
where $\nabla_\xi^2$ is the standard Laplacian for the $\xi$ coordinate, i.e.\ $\nabla_\xi^2 f = \xi^{1\!-\!D} d/d\xi(\xi^{D\!-\!1}df/d\xi)$ with $D=3$ for spheres or $D=2$ for cylinders.  The geometry of the radiation field is parameterized as $A = \nabla_\xi \cdot(\vec{F}_{\rm rad}/F_\circ)$, a dimensionless divergence.
In the two-ray approximation,
\begin{equation}
A  =  \left( - {{D\!-\!1} \over \xi}+\zeta s
\right) e^{-\tau} + \left({{D\!-\!1}\over\xi}+\zeta s \right)e^{-\tau_{\rm tot}+\tau}.
\label{eq:A}
\end{equation}

We solve the equations (\ref{eq:taux}), (\ref{eq:hydros}) and (\ref{eq:A}) subject to the boundary conditions $s = 1$ and $ds/d\xi = 0$ at the center, $\xi = 0$, \footnote{Solutions with $s(0) \neq 1$ simply correspond to a different normalization, and can be mapped onto  equivalent $s(0) = 1$ solutions.}
 and $\tau = 0$ at the outer boundary, $\xi = \xi_0$. In order to avoid matching conditions at the inner and outer boundaries, we set $\tau(0) = \tau_{\rm tot}/2$ at the center. Not all choices of $\tau_{\rm tot}$ give valid solutions, but if $\tau$ drops to zero within a finite radius then the solution is valid. We use a 4-th order Runge-Kutta integrator, and step slightly off $\xi = 0$ to avoid singularity.

\begin{figure}
  \begin{center}
    \epsfxsize = 3in
    \epsffile{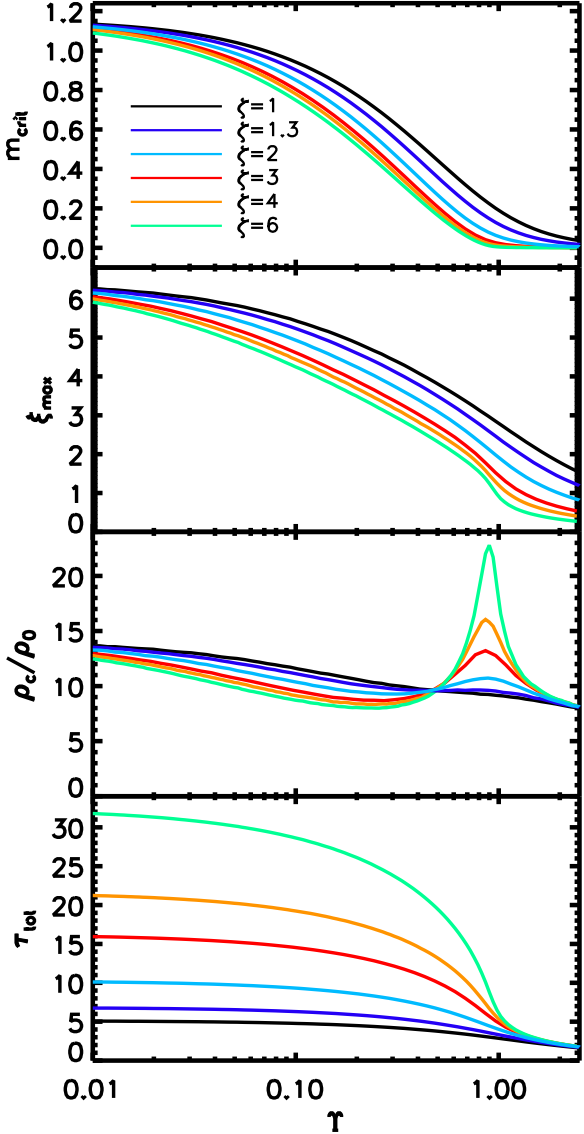}
  \end{center}
\caption{Critical values of (from top to bottom) dimensionless mass $m_{\rm crit}$, size $\xi_{\rm max}$, density contrast $\rho_{\rm c}/\rho_0$, and total optical depth $\tau_{\rm tot}$ for marginally stable irradiated spheres as a function of the dimensionless radiation pressure $\Upsilon$. Different colored curves correspond to different dimensionless opacities, $\zeta$.}
\label{core_dmless}
\end{figure}

\section{SELF-SIMILAR SOLUTIONS}\label{sec:selfsim}

\subsection{Irradiated Spheres}
The structure of irradiated, pressure-confined, isothermal spheres,  is presented in the left panel of Figure \ref{profile_core}. These dimensionless density profiles, $s(\xi)$, are calculated as described in \S\ref{sec:dimless}.  In these sample curves, a large value of optical depth, $\tau_{\rm tot}$, is assumed so that the density at the outer radius is small, i.e.\ $s(0) = s(\xi_0) < 10^{-3}$, which corresponds to a large density contrast $\rho_{\rm c}/\rho_0 = 1/s_0$.
The $\Upsilon=0$ curve corresponds to the standard Bonnor-Ebert sphere. As radiation pressure increases (to larger $\Upsilon$), the outer density profile steepens and spheres become radially truncated. Physically, the outward pressure gradient force must increase to balance the inward radiation pressure.

Gravitational stability depends on the curve of dimensionless mass, $m$, versus density contrast. In terms of the dimensional mass $M$ and surface pressure $P_0$,
\begin{equation}
m \equiv {P_0^{1/2} G^{3/2} M \over {c'_{\rm s}}^4} = \sqrt{s_0 \over 4 \pi}\int_0^{\xi_0}s\xi^2 d\xi\, .
\label{dml_m}
\end{equation}
From the right hand side above, we note that as $s_0$ decreases (and thus the density contrast increases), the prefactor decreases $\propto \sqrt{s_0}$, while the integral increases, due to the larger radius, $\xi_0$.  This mathematical competition affects gravitational stability.

Pressure bound spheres are gravitational unstable if \citep{bardeen65, stahler83}:
\begin{equation}
{\partial m \over \partial (\rho_c/\rho_0)}~ <~ 0.
\label{crit}
\end{equation}
This instability criterion is equivalent to the more intuitive Boyle's law criterion, $\partial P_0 /\partial V_0> 0$, that gravitating spheres are unstable if an enhanced surface pressure induces expansion to a larger volume, $V_0$  \citep{bonnor56, lombardi01}. In Appendix \ref{sec:Bonnor} we verify that this established correspondence also applies in the presence of other forces, such as radiation pressure.

The right panel of Figure \ref{profile_core} shows $m$ versus density contrast for different values of radiation pressure, $\Upsilon$. At low $m$ solutions are gravitationally stable since $m(\rho_{\rm c}/\rho_0)$ has a positive slope. The local maximum where $d m/d (\rho_c/\rho_0)=0$ defines marginal stability at the critical mass, $m_{\rm crit}$. For $\Upsilon = 0$, we reproduce the well known Bonnor-Ebert mass, $m_{\rm crit} = 1.18$, and the maximum density contrast of 14.1. As radiation pressure increases (larger $\Upsilon$),  both $m_{\rm crit}$  and the critical density contrast decrease. (But see below for a case where irradiation cause the critical density contrast to increase).

The properties of marginally unstable irradiated spheres are further explored in Figure \ref{core_dmless}, which also examines the effect of extinction, via $\zeta$.  All the panels in this figure correspond to the critical state with $m = m_{\rm crit}$, whose values are shown in the top panel.
Both this critical mass and the corresponding radius (shown in the second panel)  become smaller as either $\Upsilon$ or $\zeta$ increases. Either stronger irradiation or a greater opacity increases the surface radiation pressure force, which scales as $\Upsilon \zeta$ (see equation \ref{eq:hydros}).

The density contrast of critical spheres displays interesting behavior, shown in the thrid panel of Figure \ref{core_dmless}.  For small extinctions, $\zeta \lesssim 1.3$, the density contract decreases gradually with increasing irradiation, as might be expected for smaller, lower mass spheres.  For larger extinctions however, the density contrast develops a spike near $\Upsilon = 1$.  We note that our normalization of the radiation pressure (to the central pressure) is clearly appropriate since interesting behavior occurs for $\Upsilon$ near unity.

The origin of the spike in density contrast is explained by the total optical depth of the critical spheres, shown in bottom panel of  Figure \ref{core_dmless}.  The spike in the density contrast corresponds to the transition from high to low total optical depth.  The plot shows that only weakly irradiated clouds ($\Upsilon < 1$) can have a high total optical depth and remain stable.  In this weakly irradiated regime, the optical depth scales simply with the opacity, via $\zeta$, as radiative effects are not yet significantly affecting cloud structure or stability.

For stronger irradiation, as $\Upsilon$ approaches unity, the maximum optical depth of cores decreases, consistent with their smaller masses and sizes.  The spike in density contrast occurs because strong radiation pressure forces, which steepen the density profile, are being felt throughout more of the sphere.  For even stronger irradiation, with $\Upsilon$ exceeding unity, the core becomes so transparent that the radiative effects weaken, due to the flux cancellation depicted in Figure \ref{fig:two_ray}.  In this highly irradiated regime, the critical density contrast decreases again and the behavior of critical spheres is surprisingly simple: the total optical depth is order unity value for all extinction values.

In summary, as irradiation increases, the marginally gravitationally stable state gradually transitions from the Bonnor-Ebert sphere to the sphere with optical depth near unity.

\begin{figure}
  \begin{center}
    \leavevmode
    \epsfxsize = 3in
    \epsffile{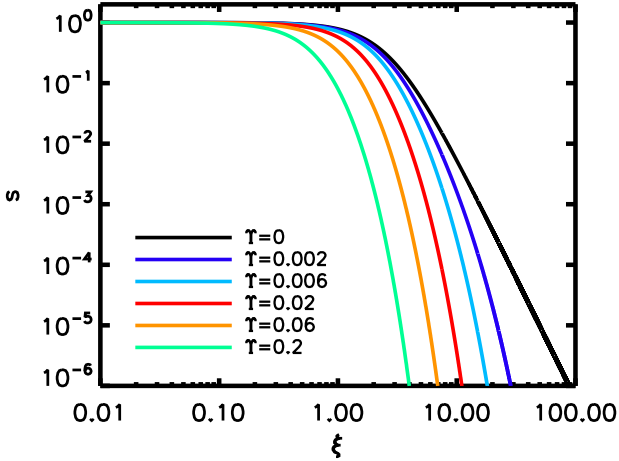}
  \end{center}
\caption{{Density profiles of dusty gas cylinders irradiated by non-ionizing radiation as a function of radius. Different colors denote different radiation pressure $\Upsilon/\zeta$. The dimensionless extinction, $\zeta$, is fixed to be 2.6. }}
\label{profile_cyl}
\end{figure}

\begin{figure}
  \begin{center}
    \leavevmode
    \epsfxsize = 3in
    \epsffile{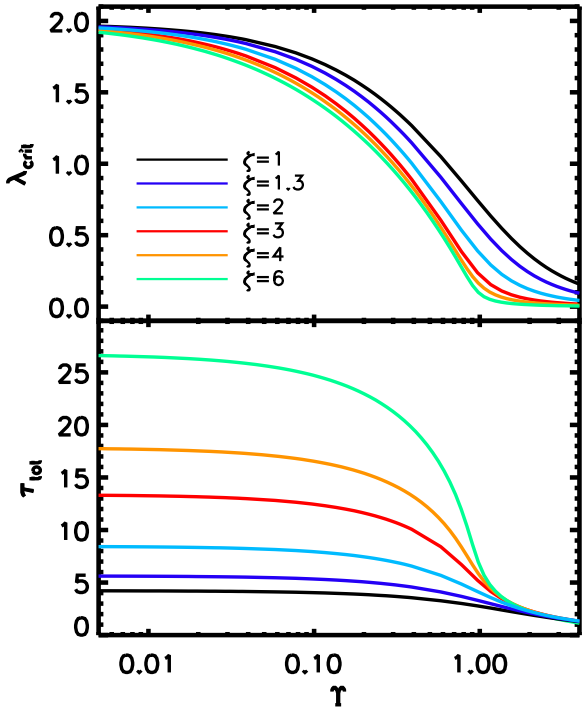}
  \end{center}
\caption{{Critical line density (top) and total optical depth (bottom) of dusty gas cylinder irradiated by non-ionizing radiation as a function of the radiation pressure $\Upsilon/\zeta$. Different colors denote different dimensionless extinction, $\zeta$.}}
\label{cyl_dmless}
\end{figure}

\subsection{Irradiated Cylinders}

Hydrostatic density profiles of irradiated cylinders are shown in Figure \ref{profile_cyl}.  Similar to the spherical case, irradiation steepens the density profile of cylinders at their outer edges.  Non-irradiated isothermal cylinder already have very steep outer density profiles, with $s \propto \xi^{-4} (\rho \propto r^{-4})$.  In the irradiated cases, even less mass resides at large radii.

An isothermal cylinder does not experience a Bonnor-type instability because $\partial P_0 /\partial V_0$ is always negative \citep{lombardi01}. Instead, the cylinder becomes unstable only when its line density exceeds the critical line density, which is the line density of a hydrostatic cylinder with  infinite outer radius. The critical line density in a dimensionless form is given as
\begin{equation}
\lambda_{\rm crit} \equiv{G \Lambda_{\rm crit} \over c_s^{'2}} = {1\over 2} \int_0^{\infty}\xi s d\xi
\label{lcrit}
\end{equation}
where $\Lambda_{\rm crit}$ is the dimensional critical line density of a hydrostatic cylinder.  With formally infinite outer radius, the critical cylinder does not have a corresponding radius or density contrast.  In reality of course cylinders are not infinite in radius, and they do have a finite density contrast set by the ambient medium.   However the steepness of the outer density profile means that the mass of the infinite clyinder is a very good approximation for the stability boundary for finite radius cylinders.  Because there is no Bonnor type instability, the external gas pressure does not affect the result.  The optical depth of the critical cylinder is well defined, again due the the steep density profile.

The critical line density and optical depth of irradiated cylinders are shown in Figure \ref{cyl_dmless}. The qualitatively behavior is  the same as the spherical case. With low levels of irradiation, $\Upsilon \ll 1$, we recover the standard result for the critical line density of isothermal cylinders, $\lambda_{\rm crit} = 2$.  Increases in either the irradiation or the opacity ($\Upsilon$ or $\zeta$) decrease the critical line density.  When radiation effects are strong, for $\Upsilon >1$, critical cylinders converge to an order unity optical depth, independent of opacity.

\begin{figure*}
  \begin{center}
    \leavevmode
    \epsfxsize = 6.5in
    \epsffile{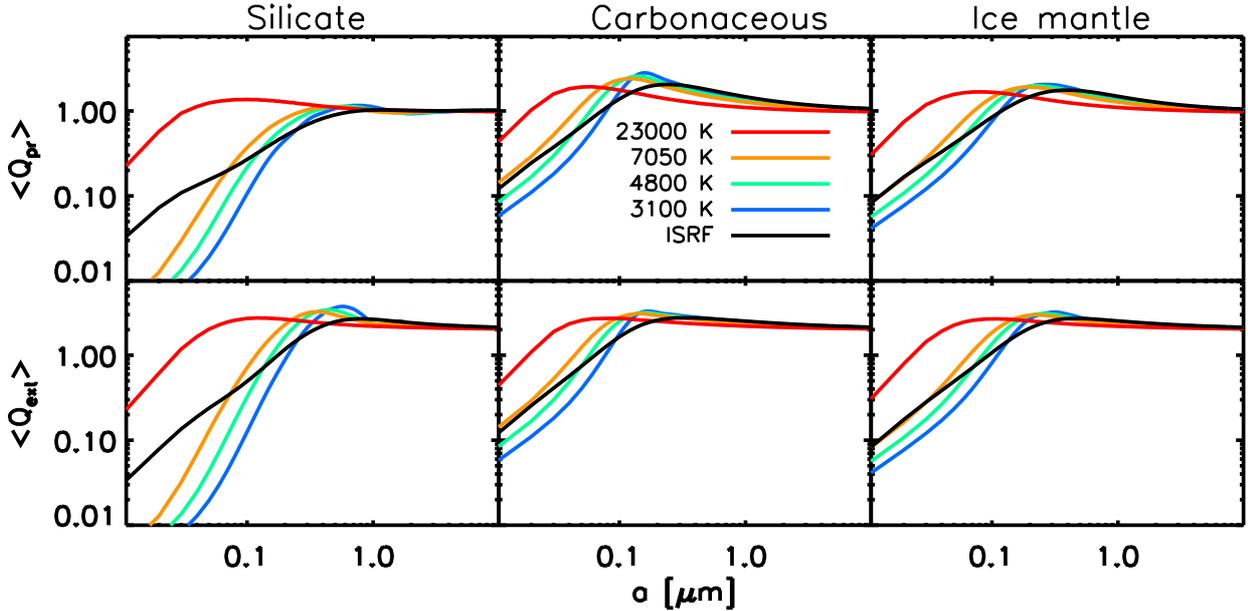}
  \end{center}
\caption{{Spectrum-averaged radiation pressure efficiency (top) and spectrum-averaged extinction (bottom) as a function of dust grain size. {\it Silicate} denotes that dust grains are composed of astro-silicates, {\it carbonaceous} denotes carbonaceous dust grains, and dust grains of {\it ice mantle} are assumed to be composed of 1:1:1 of astro-silicates, carbonaceous, and water ice.}}
\label{qpr}
\end{figure*}

\section{Results Applied to Star Forming Regions}\label{sec:applied}
We now apply our dimensionless solutions to a wide range of interstellar conditions. We summarize the adopted physical parameters in section \ref{sec:params}.  We present our numerical solutions in section \ref{sec:numsols}.   In sections \ref{sec:lowmass} and \ref{sec:highmass}, we discuss implications for low and high mass star forming regions, respectively.

\subsection{Physical Parameters} \label{sec:params}
Our solutions have a constant central number density, $n_{\rm c}$, of 10$^3$  or  10$^4$ cm$^{-3}$.  These values  characterize quasi-spherical clumps and young cores, respectively \citep{bergin07}.  Elongated filaments also span this range of central densities \citep{arzoumanian11}.

The effective sound speed
\begin{equation}
c_{\rm s} = \sqrt{0.188^2~\left({T_{\rm k} \over 10~{\rm K}}\right) + \sigma_{\rm nt}^2}~{\rm km / s} = 0.188  \sqrt{T_{\rm eff} \over 10~{\rm K}} ~{\rm km / s},
\label{csc}
\end{equation}
assumes a mean molecular weight of 2.33 proton masses. A kinetic temperature of $T_{\rm k} \sim 10$ K is a typical value for dense molecular gas  \citep[e.g.][]{Leung75,hotzel02,tafalla04}.
The non-thermal component of the velocity dispersion,$\sigma_{\rm nt}$,  provides at most an order unity correction \citep[e.g.][]{goodman98,pineda10,hacar13,seo15}.  Our numerical study considers a ragne of effective temperatures, $T_{\rm eff} = 7.5$ -- $25 ~{\rm K}$. The corresponding scale height is
\begin{equation}
 \alpha = 0.034~{\rm pc}~\left({c_{\rm s}\over 0.188{\rm km/s}}\right) \left({10^4{\rm cm^{-3}} \over n_c}\right)^{1\over2}.\label{alpha}
\end{equation}

We fix the dust-to-gas ratio to be $Z= 0.01$, a typical interstellar medium value \citep{draine07}. For grain sizes, we adopt $a$ = 0.05, 0.075, and 0.1 $\mu$m, which are characteristic sizes in molecular regions \citep{kohler15}. Though not considered here, our analysis could be extended to accommodate particle size distributions.

We use three models for the chemical composition of dust grains: (1) astro-silicate grains \citep{weingartner01}, (2) carbonaceous grains \citep{draine03} and (3) grains with an ice mantle covering  a silicate-carbonaceous core. For the ice mantle grains, we assume a 1:1:1 ratio of  astro-silicate, carbonaceous material and water ice by volume, in agreement with  \citet{li97}. The material density of dust is assumed to be 1 g/cm$^3$ (roughly 50\% porosity) for ice grains and ice mantles,  1 g/cm$^3$ for carbonaceous (about 40\% porosity) and 2 g/cm$^3$ (about 40\% porosity) for astro-silicates.

For the radiation field we use the spectrum of the ISRF in the  vicinity \citep{mezger82, mathis83}. Adopted values of the flux, $F_0$, range from $5 \times 10^{-4}$ $-$ 0.4 erg/cm$^2$/s. In the Solar vicinity (Galactocentric distance D$_G \sim$  8 kpc), the mean intensity of the non-ionizing ISRF may range from 0.015 erg/cm$^2$/s to 0.15 erg/cm$^2$/s \citep{keene81, mezger90, launhardt13}. These intensities are equivalent to a directed flux of $F_0$ $=$ 0.0025 $-$ 0.025 erg/cm$^2$/s).

We calculate the radiative efficiency parameters $\langle Q_{\rm pr}\rangle$ and $\langle Q_{\rm ext}\rangle$ with the Mie theory code, {\it miex} \citep{wolf04} for the grain and radiation properties described above. Figure \ref{qpr} shows results of the Mie calculations, both for the adopted ISRF and (for comparison) for main sequence stellar spectra of various effective temperatures (from \citealp{pickles98}).

With the above physical parameters, our dimensionless parameters scale as:
\begin{align}
\begin{split}
 \Upsilon = 0.49 &\left({\langle Q_{\rm pr}\rangle \over \langle Q_{\rm ext}\rangle}\right) \left({F_0 \over 0.02~{\rm erg/cm^2/s}}\right)\\
 &\left({10^3 {\rm cm^{-3}} \over n_{\rm c}}\right)
 \left({0.188 {\rm km/s} \over c_{\rm s}}\right)^2
\end{split}\\
\begin{split}
\zeta = 0.94&\left({Z \over 0.01}\right)\left({\langle Q_{\rm ext}\rangle \over 1}\right)\left({ n_{\rm c} \over 10^3~{\rm cm^{-3}} }\right)^{1\over2}\left({c_{\rm s} \over 0.188 ~{\rm km/s}}\right)\\
&\left({0.1~{\rm \mu m} \over a}\right)\left({1~{\rm g/cm^3} \over \rho_{\rm m}}\right)\, ,
\end{split}
\end{align}
for the radiation pressure strength and extinction, respectively.  The fact that these parameters are order unity suggests (following the self-similar analysis of section \ref{sec:selfsim}) that irradiation will indeed be important in moderately dense regions.

\begin{figure*}
  \begin{center}
    \leavevmode
    \epsfxsize = 6.5in
    \epsffile{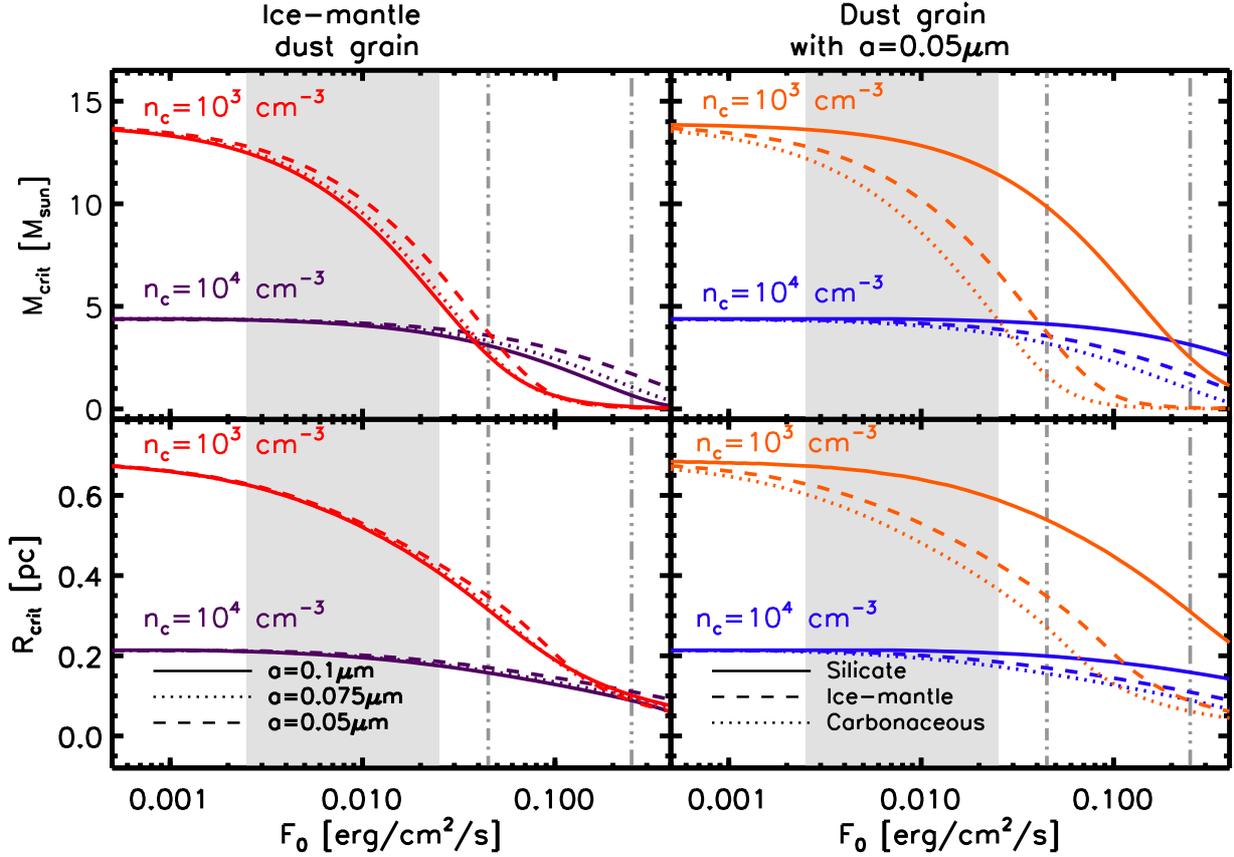}
  \end{center}
\caption{{Critical mass (top), critical size (bottom) of molecular clumps and cores as a function of radiation strength. The left panels show the critical mass and size of clumps and cores with ice-mantle dust grains. The right panels show the critical mass and size for different chemical compositions of dust grains, while dust size is fixed to be $a$ = 0.05 $\mu$m. The effective temperature is fixed to be 10 K in all solutions. The central gas densities $n_c$ are written in the panels and marked with different colors. Different line styles denote different sizes of dust grain (left panels) and different chemical composition of dust grain (right panels). The grayed area denotes a range of the ISRF strength in the Solar vicinity. The dashed-dotted and the dashed-double dotted gray lines denote the average ISRF at the molecular ring (D$_G$ = 4 kpc) and the Galactic center, respectively.}}
\label{core_dm}
\end{figure*}
\begin{figure*}
  \begin{center}
    \leavevmode
    \epsfxsize = 6.5in
    \epsffile{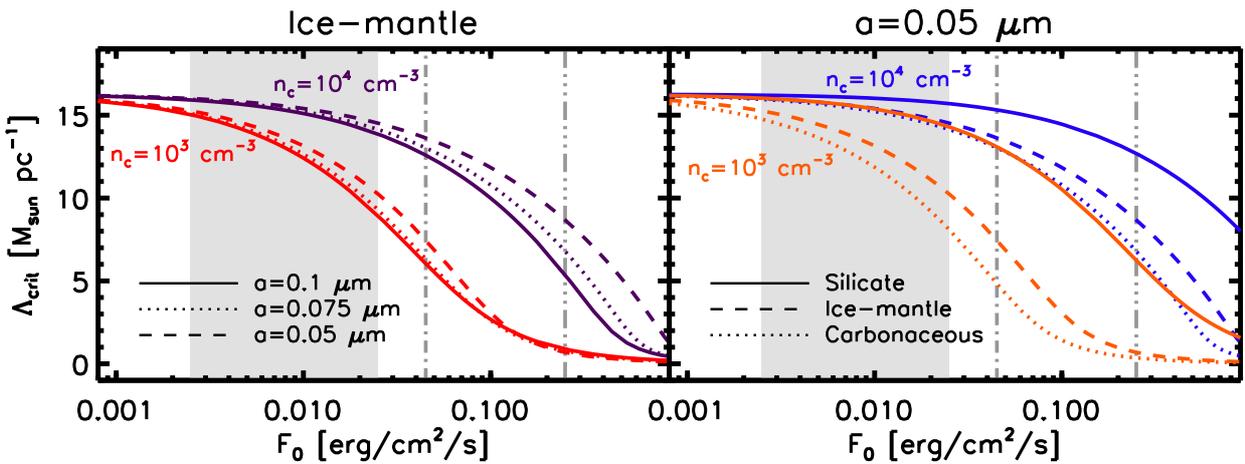}
  \end{center}
\caption{{Critical line density of molecular filaments as a function of radiation strength. Left panel shows the critical line densities with ice-mantle dust grains. Right panel show the critical line densities with different chemical composition of dust grains, while dust size is fixed to be $a$ = 0.05 $\mu$m. The effective temperature is fixed to be 10 K in all solutions. The central gas densities $n_c$ are written in the panels and marked with different colors. Different line styles denote different size of dust grain (left) and different chemical composition of dust grain (right). The grayed area denotes a range of the ISRF strength in the Solar vicinity. The dashed-dotted and the dashed-double dotted gray lines denote the average ISRF at the molecular ring (D$_G$ = 4 kpc) and the Galactic center, respectively.}}
\label{cyl_dm}
\end{figure*}

\subsection{Gravitational Instability Criteria with Irradiation}\label{sec:numsols}
\subsubsection{Molecular clumps and cores}

Figure \ref{core_dm} shows the critical mass and size of irradiated molecular clumps ($n_{\rm c}$ = 10$^3$ cm$^{-3}$) and young cores ($n_{\rm c}$ = 10$^4$ cm$^{-3}$), modeled as 10 K spheres. In the left panels, we use ice-mantle dust grains with three different dust sizes.   With weak irradiation, lower density clumps have a  higher critical mass 
and larger critical radius. 
This behavior agrees with the standard density scaling of isothermal Bonnor-Ebert spheres, $M_{\rm crit} \propto R_{\rm crit} \propto 1/\sqrt{n_{\rm c}}$.

As with the self-similar solutions, increased external irradiation decreases critical masses and radii.  A given level of irradiation more strongly affects lower density  clumps, because radiation pressure is larger relative to the central pressure.    Indeed the  irradiation parameter, $\Upsilon \propto F_0 / n_{\rm c}$ at constant temperature, quantifies this pressure ratio.  Thus while the critical state of clumps is significantly downsized by the radiation fields characteristic of the Solar neighborhood (the grey bands in Figure \ref{core_dm}), denser cores are significantly affected only when the radiation field approaches levels seen in Galactic center.

The left panels of Figure \ref{core_dm} also show that the choices of grain size only modestly affect the critical state of spheres.  This result is consistent with the fact (seen in Figure \ref{qpr}) that efficiency parameters, $\langle Q \rangle$, increase with the size of small grains so that the relevant combination $\langle Q \rangle /a$ varies weakly with size in this regime.  In other words, optical properties are more sensitive to the grain mass than the grain surface area in the long wavelength regime.

The effects of grain composition are shown in the right panels of Figure \ref{core_dm}. For silicate grains, the effects of irradiation are weaker -- as measured by the reduction in critical mass and radius at a given flux -- than for carbonaceous or ice-mantle grains.  This effect is explained by the lower radiative efficiencies of small silicate grains, as shown in Figure \ref{qpr}.   In a dense molecular region, temperatures are low enough that we do not expect bare silicates.  Either the ice-mantle case or  carbonaceous cases should be more realistic, depending on the fractional abundance of carbon species \citep{greenberg99}.

Table \ref{table1} present critical masses and radii of irradiated spheres for a range of effective temperatures.  Critical masses and radii are larger at higher temperatures.  Also, at higher temperatures, and thus higher central pressures, a correspondingly stronger external radiation pressure is needed to lower the critical masses and radii.

\subsubsection{Molecular filaments}

The critical line density of irradiated filaments at 10 K are shown in Figure \ref{cyl_dm}. In the low-irradiation limit, the critical line density, $\Lambda_{\rm crit} = 2 c_{\rm s}^2/G$ is independent of central density, unlike the $M_{\rm crit}$ of spheres. Analogously to the spherical case, the critical line density of cylinders decreases for increasing levels of irradiation. Also like the spherical case, irradiation is felt more strongly by lower density cylinders, grain size dependence is weak and silicate grains are less affected by irradiation. The different geometry of spheres and cylinders gives rise to some subtle differences. For instance $\Lambda_{\rm crit}$ is slightly slightly less sensitive to increasing irradiation than is $M_{\rm crit}$. Table \ref{table2} presents critical line densities of irradiated cylinders for a range of effective temperatures.

\begin{table*}
	\centering
	\caption{Critical Sizes and Masses of Dense Clumps and Cores under Radiation}
	\label{table1}
	\begin{tabular}{ c c c c c c c c c } 
		\hline
  $n_c^a$      & $F_0^b$        &  $a^c$     & &  \multicolumn{4}{|c|}{Critical Size \& Mass   (pc, M$_\odot$) }   \\  \cline{5-9}
  (cm$^{-3}$)  &(erg/cm$^2$/s)  &  ($\mu$m)   & & $T_{\rm eff}^d$ = 7.5 K & 10 K & 15 K & 20 K & 25 K  \\
  \hline
      &   0.002   & 0.1    & & 0.546   , 8.08   &  0.636   ,12.6    &  0.789   ,23.7    &  0.919   ,36.9    &  1.03    ,52.0    \\
      &           & 0.05   & & 0.547   , 8.25   &  0.639   ,12.9    &  0.742   ,22.5    &  0.924   ,37.5    &  1.04    ,52.7    \\
      &   0.006   & 0.1    & & 0.481   , 6.66   &  0.568   ,10.7    &  0.716   ,20.8    &  0.842   ,33.2    &  0.954   ,47.4    \\
      &           & 0.05   & & 0.486   , 7.08   &  0.574   ,11.4    &  0.633   ,18.3    &  0.853   ,34.7    &  0.967   ,49.4    \\
      &   0.02    & 0.1    & & 0.360   , 3.55   &  0.438   , 6.25   &  0.572   ,13.7    &  0.688   ,23.4    &  0.792   ,35.2    \\
      &           & 0.05   & & 0.375   , 4.29   &  0.455   , 7.60   &  0.461   , 9.17   &  0.712   ,27.3    &  0.819   ,40.4    \\
10$^3$&   0.06    & 0.1    & & 0.212   , 0.853  &  0.267   , 1.63   &  0.374   , 4.32   &  0.477   , 8.84   &  0.570   ,15.3    \\
      &           & 0.05   & & 0.230   , 0.978  &  0.303   , 2.32   &  0.231   , 0.926  &  0.528   ,14.5    &  0.622   ,23.9    \\
      &   0.2     & 0.1    & & 0.0968  , 0.0977 &  0.118   , 0.171  &  0.159   , 0.396  &  0.201   , 0.76   &  0.246   , 1.32   \\
      &           & 0.05   & & 0.0825  , 0.0552 &  0.106   , 0.109  &  0.0722  , 0.0381 &  0.235   , 0.941  &  0.327   , 2.56   \\
      &   0.6     & 0.1    & & 0.0498  , 0.0145 &  0.0590  , 0.0239 &  0.0758  , 0.0491 &  0.0913  , 0.0837 &  0.106   , 0.129  \\
      &           & 0.05   & & 0.0384  , 0.00640&  0.0464  , 0.0109 &  0.0343  , 0.0046 &  0.0764  , 0.0448 &  0.0921  , 0.0748 \\
\hline
      &   0.002   & 0.1    & &   0.183   , 2.77   &  0.212   , 4.29   &  0.261   , 7.91   &  0.303   ,12.2    &  0.339   ,17.1    \\
      &           & 0.05   & &   0.184   , 2.79   &  0.213   , 4.31   &  0.262   , 7.95   &  0.304   ,12.3    &  0.340   ,17.2    \\
      &   0.006   & 0.1    & &   0.174   , 2.66   &  0.204   , 4.14   &  0.253   , 7.72   &  0.294   ,12.0    &  0.331   ,16.8    \\
      &           & 0.05   & &   0.177   , 2.72   &  0.207   , 4.22   &  0.256   , 7.83   &  0.298   ,12.1    &  0.335   ,17.0    \\
      &   0.02    & 0.1    & &   0.154   , 2.32   &  0.183   , 3.72   &  0.231   , 7.12   &  0.272   ,11.2    &  0.309   ,15.9    \\
      &           & 0.05   & &   0.160   , 2.49   &  0.189   , 3.94   &  0.239   , 7.46   &  0.281   ,11.7    &  0.319   ,16.5    \\
10$^4$&   0.06    & 0.1    & &   0.124   , 1.63   &  0.150   , 2.81   &  0.195   , 5.80   &  0.234   , 9.53   &  0.270   ,13.9    \\
      &           & 0.05   & &   0.133   , 2.01   &  0.161   , 3.32   &  0.209   , 6.58   &  0.250   ,10.6    &  0.287   ,15.2    \\
      &   0.2     & 0.1    & &   0.0798  , 0.464  &  0.102   , 1.09   &  0.141   , 3.04   &  0.175   , 5.76   &  0.206   , 9.14   \\
      &           & 0.05   & &   0.0944  , 1.030  &  0.118   , 1.97   &  0.161   , 4.55   &  0.198   , 7.89   &  0.232   ,11.9    \\
      &   0.6     & 0.1    & &   0.0263  , 0.0159 &  0.0373  , 0.0397 &  0.0732  , 0.271  &  0.106   , 1.17   &  0.133   , 2.74   \\
      &           & 0.05   & &   0.0472  , 0.0725 &  0.0694  , 0.383  &  0.106   , 1.68   &  0.138   , 3.76   &  0.167   , 6.51   \\
		\hline
\multicolumn{9}{l}{\textsuperscript{a}\footnotesize{The central density of gas}}\\
\multicolumn{9}{l}{\textsuperscript{b}\footnotesize{The radiation flux normal to the dense clump/core surface}}\\
\multicolumn{9}{l}{\textsuperscript{c}\footnotesize{The radius of dust grain. The ice-mantle dust grain model is used.}}\\
\multicolumn{9}{l}{\textsuperscript{d}\footnotesize{The effective temperature of the internal supports including thermal and non-thermal components}}
	\end{tabular}
\end{table*}
\begin{table*}
	\centering
	\caption{Critical Line Densities of Filaments under Radiation}
	\label{table2}
	\begin{tabular}{ c c c c c c c c c } 
  \hline
  $n_c^a$      & $F_0^b$        &  $a^c$  &    & \multicolumn{5}{|c|}{  Critical Line Density (M$_\odot$/pc)}   \\
  \cline{4-9}
  (cm$^{-3}$)  &(erg/cm$^2$/s)  &  ($\mu$m)   &  $T_{\rm eff}^d$ = & 7.5 K & 10 K & 15 K & 20 K & 25 K  \\
  \hline
      &   0.002   & 0.1    & &  11.4    & 15.3    & 23.2    & 31.2    & 39.2  \\
      &           & 0.05   & &  11.5    & 15.4    & 23.4    & 31.4    & 39.4  \\
      &   0.006   & 0.1    & &  11.2    & 15.2    & 23.0    & 31.0    & 38.9  \\
      &           & 0.05   & &  10.9    & 14.5    & 22.3    & 30.1    & 38.0  \\
      &   0.02    & 0.1    & &  10.1    & 13.8    & 21.4    & 29.1    & 36.8  \\
      &           & 0.05   & &  10.4    & 14.2    & 21.9    & 29.7    & 37.5  \\
10$^3$&   0.06    & 0.1    & &   9.82   & 13.5    & 20.9    & 28.5    & 36.2  \\
      &           & 0.05   & &   9.11   & 12.2    & 19.4    & 26.8    & 34.4  \\
      &   0.2     & 0.1    & &   7.17   & 10.2    & 16.7    & 23.6    & 30.6  \\
      &           & 0.05   & &   7.72   & 11.1    & 18.1    & 25.3    & 32.7  \\
      &   0.6     & 0.1    & &   6.58   &  9.46   & 15.7    & 22.4    & 29.2  \\
      &           & 0.05   & &   5.30   &  7.07   & 13.0    & 19.5    & 26.1  \\
\hline
      &   0.002   & 0.1    & &  12.0    & 16.0    & 24.1    & 32.2    & 40.3  \\
      &           & 0.05   & &  12.0    & 16.1    & 24.1    & 32.3    & 40.4  \\
      &   0.006   & 0.1    & &  11.9    & 16.0    & 24.0    & 32.1    & 40.2  \\
      &           & 0.05   & &  11.8    & 15.8    & 23.8    & 31.9    & 40.0  \\
      &   0.02    & 0.1    & &  11.6    & 15.6    & 23.6    & 31.6    & 39.6  \\
      &           & 0.05   & &  11.7    & 15.7    & 23.7    & 31.8    & 39.8  \\
10$^4$&   0.06    & 0.1    & &  11.5    & 15.4    & 23.4    & 31.4    & 39.4  \\
      &           & 0.05   & &  11.2    & 15.1    & 23.0    & 31.0    & 39.0  \\
      &   0.2     & 0.1    & &  10.5    & 14.3    & 22.1    & 29.9    & 37.8  \\
      &           & 0.05   & &  10.9    & 14.8    & 22.6    & 30.5    & 38.5  \\
      &   0.6     & 0.1    & &  10.3    & 14.1    & 21.8    & 29.5    & 37.4  \\
      &           & 0.05   & &   9.77   & 13.5    & 21.1    & 28.7    & 36.5  \\
		\hline
\multicolumn{9}{l}{\textsuperscript{a}\footnotesize{The central density of gas}}\\
\multicolumn{9}{l}{\textsuperscript{b}\footnotesize{The radiation flux normal to the filament surface}}\\
\multicolumn{9}{l}{\textsuperscript{c}\footnotesize{The radius of dust grain. The ice-mantle dust grain model is used.}}\\
\multicolumn{9}{l}{\textsuperscript{d}\footnotesize{The effective temperature of thermal and non-thermal support.}}
	\end{tabular}
\end{table*}

\subsection{Low-Mass Star Forming Regions}\label{sec:lowmass}

In low-mass star forming regions, both nearby young stellar objects (YSOs) and the diffuse ISRF contribute to the radiation field. We first consider quiescent regions with no YSOs.  Under the ISRF in the Solar vicinity (shaded gray in Figure \ref{core_dm}), the critical mass of a molecular clump ($n_{\rm c}$ = 10$^3$ cm$^{-3}$) is significantly reduced, to values as small as 5.5 M$_\odot$, i.e.\ 40\% of the non-irradiated value (see red curves in of Figure \ref{core_dm}). On the other hand, the stability properties of  dense cores ($n_c$ $>$ 10$^4$ cm$^{-3}$) are only modestly affected by the Solar ISRF  (see blue curves in Figure \ref{core_dm}). Thus while the Solar ISRF does not affect the collapse of dense cores, this level of irradiation can affect the formation of dense cores within less dense clumps.

Closer to the Galactic center, the ISRF is more intense, around $\sim$9 times stronger in the molecular ring (D$_G$ = 4 kpc) compared to the Solar vicinity, and $\sim$50 times stronger at the Galactic center \citep{mezger90}.  (These levels of irradiation are marked by  dashed-dotted and dashed-triple dotted lines, respectively, in Figure \ref{core_dm}.)
In the inner Galactic regions, the critical sizes and masses of both dense clumps and dense cores may be significantly smaller ($<$20\%) than those in the Solar vicinity.  At face value, these results favor the formation of  lower mass stars in the more intensely irradiated inner Galaxy.  On the other hand, competing effects -- such as higher temperatures \citep[e.g.][]{walmsley83, ao13} and extra non-thermal (e.g.\ magnetic) support -- favor higher mass star formation towards the Galactic center.   More detailed models are needed to include these effects self-consistently.

In more evolved regions, molecular clouds recieve enhanced irradiation from nearby YSOs, stellar associations or clusters (e.g. NGC 1333, B1688 \& B1689 in $\rho$ Ophiuchus, L1495A in Taurus, see \citealp{volgenau06, bontemps01, maruta10, seo15}). Consider a stellar association following  Kroupa's IMF (initial mass function) with $\sim$600 stars and with  the earliest stellar type being B9V.    For a typical cluster diameter of 1 pc \citep{nilakshi02}, we roughly and conservatively estimate the radiation field by placing all stars in a 0.5 pc radius shell around a central reference point.  A non-ionizing radiation flux of  0.05 erg/cm$^2$/s would be incident upon an opaque cloud at this reference point.  The actual radiation field will of course depend on more realistic stellar locations and will be larger closer to the brightest stars.  Figure \ref{core_dm} shows that this nominal flux of  0.05 erg/cm$^2$/s is enough to reduce the critical mass of dense cores from 4.4 $M_\odot$ to $\sim 3 M_\odot$.  This  order unity change might not seem significant, given our idealized model.  However the idea that radiation from nearby YSOs could help regulate the IMF merits further discussion, which we begin in the summary.

Irradiation also helps to trigger the formation of starless cores  within filamentary structures.
A non-irradiated young filament with $n_{\rm c}$ $\sim$ 10$^3$ cm$^{-3}$ at 10 K should not fragment until it reaches a critical line density of 16.4 M$_\odot/$pc.  However irradiation by the ISRF in the Solar vicinity can lower the fragmentation threshold to 9 M$_\odot$/pc  (see gray shaded area in the left panel of Figure \ref{cyl_dm}).  Irradiation can thus explain why some filaments embed dense cores, despite having line densities below the standard (non-irradiated) critical value \citep{benedettini15}. For example, a cold filament in L1495-B218  (\#28 in \citealp{hacar13}) has a line density of 9.3 M$_\odot/$pc but embeds NH$_3$ starless cores \citep{seo15}.   On the other hand,  filaments with high line densities, e.g. in Aquila \citep{arzoumanian11}, seem to require extra internal support -- e.g.\ magnetic --  to explain their existence beyond nominal stability thresholds.
Thus a detailed understanding of the physical conditions of a filament -- including the radiation environment -- is required to determine gravitational stability.

\subsection{High-Mass Star Forming Regions}\label{sec:highmass}

High mass star forming regions are strongly affected by radiation from OB associations. Observations preferentially find YSOs near the rims of HII regions and near globules within HII regions \citep{jose13}.  Massive stars can trigger star formation in different ways.   In the collect-and-collapse scenario,
the stellar winds of massive stars  are the trigger \citep{elmegreen77, whitworth94, hosokawa05, dale07}.    In the radiation driven implosion (RDI) scenario, radiation pressure from  ionizing photons is responsible  \citep{bertoldi89, lefloch94, white97, kessel-Deynet03, gritschneder09, chauhan09, bisbas11, walch13}.

However, triggered star formation is harder to understand far from the direct influence of  stellar winds and ionizing photons.  For instance, enhanced YSO populations are  observed deep within globules in the HII regions G028.83-0.25 and G041.92+0.04 \citep{dirienzo12}.   Large scale compression could play a role.  Alternately, we propose  that the prodigious non-ionizing radiation from OB associations helps trigger star formation in more embedded regions.  Due to a greater penetration depth, non-ionizing radiation can influence embedded regions (quantitatively N$_{\rm H}$ $\approx$ 10$^{18}$ cm$^{-2}$ for $\tau$ = 1 at  the Lyman limit of 912{\AA}, while N$_{\rm H}$ $\approx$ 10$^{21}$ cm$^{-2}$ for a visual extinction A$_{\rm V}$ = 1, \citealp{cardelli89,gay12}).  Moreover, triggered YSOs can subsequently promote sequential star formation via non-ionizing radiation pressure that extends into embedded regions.

Consider, for example, the Elephant's trunk nebula (IC1396A), often cited as an example of triggered star formation. IC1396A is a globule within an HII region containing  more than 50 embedded YSOs \citep{reach04, sicilia-aguilar05, sicilia-aguilar06a,sicilia-aguilar06b, sicilia-aguilar14, morales-calderon09, getman12} and a total mass of $\sim 200 M_\odot$ \citep{morgan10}. YSOs at the bright rim of the globule -- illuminated by an O6.5 star at 4.5 pc -- are believed to be formed through the RDI \citep{sicilia-aguilar14}. We propose that YSOs in the interior of the globule were triggered by non-ionizing radiation.  In particular many YSOs  formed in-between the bright outer rim of IC1396A and its central cavity, which is being cleared by a bright stellar association that includes the intermediate mass variable V390 Cep.   With irradiation from both sides,  the YSOs embedded in this in-between region start to resemble our idealized model of isotropic irradiation.  The flux of the OB association at the surface of globule is 0.4 erg/cm$^2$/s and the flux from the YSOs at the globule center may be 0.05 erg/cm$^2$/s if we assume that the combined spectral type of YSOs are a single A0 type star. From figure \ref{core_dm} these roughly estimated flux levels can strongly affect star formation.  While our simple model can not make detailed predictions for such a complicated environment, we hope to motivate more study of the effects of non-ionizing radiation pressure environments like IC1396A.

When HII regions are illuminated by later type stars, i.e.\  OB associations with  spectral types later than O6V, non-ionizing radiation will have a greater relative importance compared to ionizing radiation.  For instance in the Gum nebula, cometary globules such as the CG30/31 complex \citep{nielsen98, kim05} and the  CG4/Sa101 complex \citep{rebull11} are far from  OB associations, but they are near to a stellar association with multiple B and A type stars.
Similarly, the Cone nebula is a cometary dense cloud embedded within an HII region that is illuminated by Allen's source (i.e.\ the B star NGS2264IRS  \citealp{thompson98}). The YSOs in these regions are assumed to be a result of triggered star formation, even though the effects of stellar winds and ionizing radiation are relatively weak.  On the other hand, we find that the non-ionizing radiation from nearby stellar associations or late-type OB associations is strong enough to trigger gravitational collapse by reducing the critical core mass (Figures \ref{core_dm} and \ref{cyl_dm}).

\section{Model Assumptions and Extensions}

Our study focuses on the role of non-ionizing radiation pressure on dust grains in molecular filaments and cores.  Our neglect of ionizing radiation is appropriate not only for low mass star forming regions but also where ionizing sources are strongly extincted. Similarly, we neglect some radiative forces on dust grains, specifically the photoelectric and photodesorption forces \citep{weingartner01}.  These forces -- which arise when UV photons remove electrons or atoms, respectively, from the surface of a dust grain --  are more important in the diffuse interstellar medium, e.g. in the cold neutral medium, than in the dense molecular regions we consider.

We assume an isotropic background radiation field, where the anisotropy required for a radiation pressure force is introduced by an opaque cylinder or sphere.  In reality,  the background radiation will be anisotropic, though the ISRF in the Solar vicinity is only asymmetric at the 10\% level \citep{weingartner01}. For  highly  anisotropic irradiation from a nearby star or association, our results will not apply directly. However, anisotropic non-ionizing irradiation may still promote star formation in a similar way as ionizing irradiation from an OB association can trigger star formation \citep{bisbas11}. By considering the isotropic component of radiation fields here, future studies can determine the effects of more complex irradiation.


Observed cores and filaments are not the perfect spheres or cylinders used in our study. Observed dense cores are spheriods with a typical axis ratio of $\leq$2:1 \citep{myers91}. The BE stability of general spheroids is similar to that of perfect spheres. In particular, \citet{lombardi01} found that the maximum density contrast of a stable, isothermal cloud is independent of shape. While filaments are not perfect cylinders, their length-to-width ratios are typically $>$10:1 \citep{andre10, andre14, menshchikov10, hacar13}. Our idealized geometries are consistent with the isotropic irradiation and provide a reference for more complex geometries.

\begin{figure}
  \begin{center}
    \leavevmode
    \epsfxsize = 3in
    \epsffile{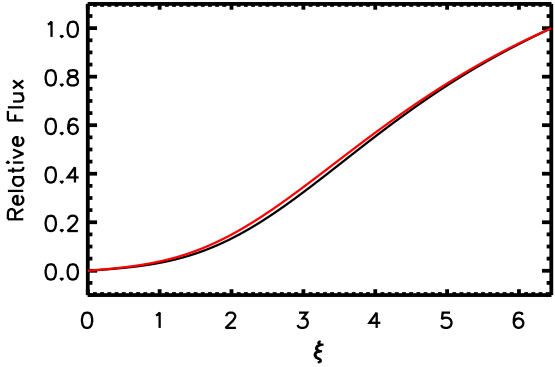}
  \end{center}
\caption{{Comparison of a two-ray approximation to a full radiative transfer calculation. A flux of radiation field is shown as a function of radius within the critical Bonnor-Ebert sphere. Total extinction, A$_{\rm V}$, is 10 for this example. Only absorption by dust grains is considered while scattering is neglected. The black line denotes a full radiative transfer calculation, and the red line denotes a two-ray approximation. The difference between the two fluxes within the sphere is $<$3\% of the flux at the sphere surface.}}
\label{asym}
\end{figure}

Our two-ray approximation to the extinction of radiation simplifies the detailed angular dependence of the radiation field, i.e. we neglect rays that obliquely intersect the radius vector.   In reality, a radiation field that is isotropic for all incoming angles at the surface would become more radially directed with depth (due to the increased slant optical depth of oblique rays).  Figure  \ref{asym} tests our approximation by comparing the two-ray approximation to a detailed extinction calculation that includes the angular dependences.  The detailed calculation shows only a minor reduction in flux, of at most a few percent.  Thus the two-ray approximation is acceptable, especially in the context of other, more severe, approximations.

The  isothermal structure in our models is only roughly consistent with observations, which find a factor of two variation in temperature within filaments and dense cores.  Central temperatures ($\sim$8 K) are typically lower than outer regions ($\sim$20 K) due to interstellar radiation heating \citep{evans01, zucconi01, ward-thompson02, pagani04, shirley05, crapsi07, launhardt13, palmeirim13, seo15}.
With a more realistic temperature structure, the critical line density of filaments increases by 20 -- 30\% \citep{recchi13}.  Non-isothermal spheres show a similar correction \citet{Sipila11}.   A more complex model with detailed heating and cooling could self-consistently determine both the temperature structure and the radiation pressure profile for a given level of irradiation.

We neglect rotation, which is probably a weak dynamical effect.  Observations of velocity variations across a filament or a dense core indicate that rotational energy is typically less than 5\% of the internal or gravitational energy \citep[e.g.][]{arquilla86, goodman93, caselli02b}.

\begin{figure}
  \begin{center}
    \leavevmode
    \epsfxsize = 3in
    \epsffile{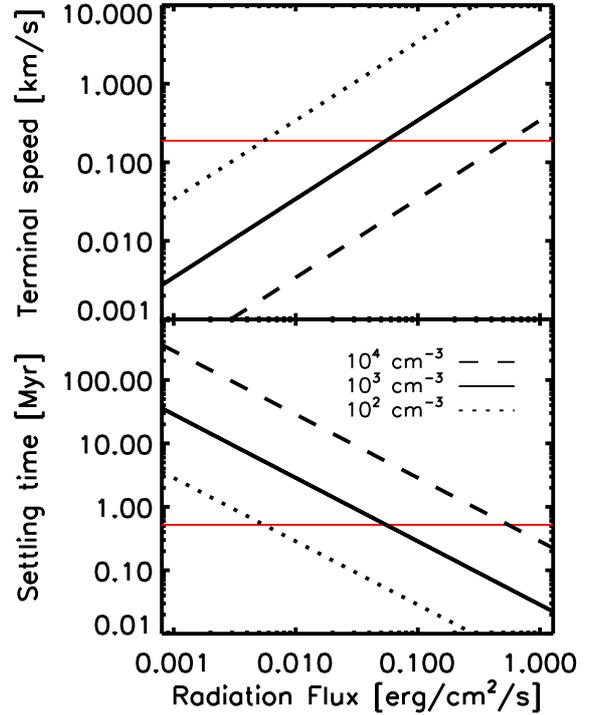}
  \end{center}
\caption{Dust grains are accelerated by radiation pressure (with flux levels on the x-axis) and these motions are damped by gas drag (with curves for different gas densities as labeled).  Plotted are both the terminal speed (top) and settling time across 0.1 pc  (bottom). Horizontal  reference lines (in red) show the sound speed (for $T =  10$ K)  and sound crossing time (over  0.1 pc).  For high fluxes and low gas densities, short settling times allow grain sedimentation.  However, internal motions could counteract sedimentation.}
\label{dset}
\end{figure}

We assume perfect coupling between dust grains and gas, neglecting grain sedimentation. Since radiation pressure dominates gravity at the surface of our objects (for most adopted irradiation levels), we estimate the terminal speed of dust grains as
\begin{equation}
v_{\rm terminal}~=~{3\langle Q_{\rm pr}\rangle F_0 \over 4  \rho_{\rm g} c_s c},
\label{vtrm}
\end{equation}
using the Epstein drag law appropriate for dilute gases \citep{houches10}.  The terminal velocity is independent of grain size since radiation pressure  and drag forces both scale with the grain cross section.  Figure \ref{dset} shows the terminal velocity and the settling time (across a typical scale of 0.1 pc) versus level of irradiation. The red lines in the plots correspond to the sound speed and sound crossing time.  With high levels of irradiation and low gas densities (appropriate near the surface) significant sedimentation could occur.  This tendency to settle is counteracted by mixing due to turbulent diffusion or other large scale motions \citep{yl07}.  Sedimentation will also be reduced for more optically thin objects that experience flux cancellation. Sedimentation of grains could alter dynamical stability by piling dust grains and concentrating the radiation force in a thin shell, and could also affect observed extinction profiles \citep{whitworth02}.  Time-dependent numerical simulations are likely required to explore the full effects of grain sedimentation and its dynamical effect on dense structures.

\section{SUMMARY \& CONCLUSIONS}

We study the gravitational stability of hydrostatic cylinders and spheres bathed in an  isotropic, non-ionizing radiation field.
We find that the radially-inward radiation pressure force promotes -- and can trigger -- the gravitational collapse of  cores and filaments.  The classic stability thresholds --  the critical mass of Bonnor-Ebert spheres and the critical line density of isothermal Ostriker cylinders -- are significantly lowered once the surface radiation pressure reaches the magnitude of the central gas pressure.  Thus a given level of irradiation more strongly affects  objects with a lower central density, provided the column density is large enough that the optical depth is at least an order unity.  Our analysis shows that the critical state of highly irradiated spheres or filaments is characterized by an order unity optical depth.  Physically, these highly irradiated objects must become partially transparent to incident irradiation to avoid implosion.

Standard interstellar radiation fields are strong enough to influence gravitational stability.  For instance, consider a spherical molecular clump with a central density of $\sim$10$^3$ cm$^{-3}$ at 10 K.  The maximum stable mass of such an object is 14 M$_\odot$ without radiation (the Bonnor-Ebert mass) and only 5 M$_\odot$ if subject to interstellar irradiation of the Solar vicinity.  For more evolved, i.e. denser, cores with central densities  $\sim$10$^4$ cm$^{-3}$ a stronger radiation field is needed to affect gravitational stability.  These stronger radiation fields can be found towards the Galactic center or in active star forming regions near YSOs or stellar associations.

The gravitational stability of molecular filaments is similarly affected by irradiation.  A young filament with a central density of $\sim$10$^3$ cm$^{-3}$ at 10 K has a maximum line density of 16 M$_\odot$/pc without radiation, the Ostriker value.  Interstellar irradiation in the Solar vicinity can lower the critical line density to only 9 M$_\odot$/pc.

At face value, our results imply that the initial mass function (IMF) of stars could vary with the radiation environment.  Alternatively, we propose a mechanism to maintain a universal IMF with non-ionizing radiation pressure.  Imagine that in some star forming region there is a temporary over- (or under-) production of massive stars, for either physical or stochastic reasons.  Our results show that the resulting enhancement (or reduction, respectively) in irradiation can then reduce (or increase, respectively) subsequent fragmentation masses.  The triggering of star formation by the evolving radiation pressure  could thus help regulate the IMF.

The full implications of non-ionizing radiation pressure are not yet clear due to neglected effects in our model.  Additional support from magnetic fields or strong turbulence will counteract the destabilizing effects of radiation pressure.
 Furthermore, our static models neglect crucial evolutionary and dynamical effects, notably the processes that set the mass spectrum of prestellar cores.  Nevertheless, our results demonstrate that non-ionizing radiation pressure is strong enough to influence both star formation and filamentary structure, so its full implications should be explored in future work.

\section*{Acknowledgements}

We are grateful to J. Serena Kim, Yancy L. Shirley and Kaitlin M. Kratter for stimulating discussions.  We thank Bruce Draine for encouragement.





\bibliographystyle{mnras}
\bibliography{library}



\appendix

\section{The Dimensionless Bonnor Stability Criterion with Radiation Pressure}\label{sec:Bonnor}

We now show that the dimensionless stability criteria of equation (\ref{crit}) applies even in the presence of radiation pressure, or for that matter any additional spherically symmetric force.  The existence of a stability boundary at  $dm/ds_0 = 0$ can be derived, for instance, from an analysis of the free energy of the system \citep{stahler83}.  We choose a simpler path and show that the dimensional Bonner instability condition in physical units, $dP_0/dV_0 > 0$ at fixed mass, 
remains equivalent to $dm/ds_0 >0$ and thus equation (\ref{crit}).

First, we express the dimensional outer radius, $r_0 = r(s_0) = \alpha \xi_0$, in terms of the dimensionless mass, $m$,  as
\begin{equation}\label{eq:r0}
r_0 = {GM \over {c'}_{\rm s}^2}  {\sqrt{s_0} \xi_0 \over m}
\end{equation}
using equation (\ref{dml_m}).

Next we consider perturbations to hydrostatic solutions at fixed dimensional mass, $M$, and temperature, i.e. $c'_{\rm s}$.  Equations (\ref{dml_m}) and (\ref{eq:r0}) give
\begin{subeqnarray}
d\ln(P_0) &=&  2 d \ln (m), \\
d\ln(r_0) &=& d\ln(\xi_0) + \tfrac{1}{2}d\ln(s_0) - d\ln(m) .
\end{subeqnarray}

We can now relate the Bonnor stability criterion to dimensionless variables as
\begin{equation}
{d\ln(P_0) \over d \ln(V_0)} = 3{d \ln(P_0) \over d \ln(r_0)} = 6{d \ln(m)/d\ln(s_0) \over {d\ln(\xi_0) \over d\ln(s_0)} +{1 \over 2} - {d\ln(m) \over d\ln(s_0)}}\, . \label{crit_comp}
\end{equation}

Thus a stability transition at $dP_0/dV_0 = 0$ also gives a transition at $dm/ds_0 = 0$.  The desired sign of the stability criterion for $dm/ds_0$ requires that the denominator in the final term of equation (\ref{crit_comp}) be positive.

We verified numerically that this denominator remains positive for our solutions, but a more general argument proceeds as follows.  First, note that this denominator is positive for the classic Bonnor-Ebert solution, as it must be since the correspondence between the stability criteria holds in this case.  Next, consider that any finite radiation pressure (or other force) can be increased incrementally from zero to produce a continuous family of solutions that starts with the Bonnor-Ebert solution (as visualized in Figure \ref{profile_core}).  Since none of these incremental steps can produce an infinite divergence in $dP_0/dr_0$, the denominator in question cannot change sign, and the desired correspondence between dimensionless and dimensional stability criteria holds.


\bsp	
\label{lastpage}
\end{document}